\documentclass[galaxies,conferencereport,submit,moreauthors,pdftex,10pt,a4paper]{mdpi} 

\firstpage{1} 
\articlenumber{3}
\pubvolume{7}
\pubyear{2019}
\copyrightyear{2019}
\history{Received: 20 November 2018; Accepted: 21 December 2018; Published: 25 December 2018}
\usepackage{subfig}

\def\apj{\rm ApJ}
\def\apjl{\rm ApJL}
\def\apjs{\rm ApJS}

\def\mnras{\rm MNRAS}

\def\nat{\rm Nature}

\def\aap{\rm A\&A}

\def\sovast{\rm SvA}
 \theoremstyle{mdpi}
 \newcounter{thm}
 \newcounter{ex}
 \newcounter{re}
\usepackage[normalem]{ulem}

\Title{ON THE CONNECTION OF RADIO AND \texorpdfstring{$\gamma$}{}-RAY EMISSION IN BLAZARS}

\Author{Stella Boula \texorpdfstring{$^{1*}$}{}, Maria Petropoulou \texorpdfstring{$^{2\dagger}$}{} and Apostolos Mastichiadis \texorpdfstring{$^{1}$}{}} 
\AuthorNames{Stella Boula, Maria Petropoulou and Apostolos Mastichiadis}

\address{%
$^{1}$ Department of Physics, National and Kapodistrian University of Athens\\
$^{2}$ Department of Astrophysical Sciences, Princeton University \\
$^{\dagger}$ Lyman Spitzer Postdoctoral Fellow}
\corres{Correspondence: stboula@phys.uoa.gr}

\abstract{Blazars are a sub-category of radio-loud active galactic nuclei with relativistic jets pointing towards to the observer. They are well-known for their non-thermal variable emission, which  practically extends over the whole electromagnetic spectrum. Despite the plethora of multi-wavelength observations, the issue about the origin of the \texorpdfstring{$\gamma$}{}-ray and radio emission in blazar jets remains unsettled. Here, we construct a parametric leptonic model for studying the connection between the \texorpdfstring{$\gamma$}{}-ray and radio emission in both steady-state and flaring states of blazars. Assuming that relativistic electrons are injected continuously at a fixed distance from the black hole, we numerically study the evolution of their population as it propagates to larger distances while losing energy due to expansion and radiative cooling. In this framework, \texorpdfstring{$\gamma$}{}-ray photons are naturally produced at small distances (e.g. \texorpdfstring{$10^{-3}$}{} pc) when the electrons are still very energetic, whereas the radio emission is produced at larger distances (e.g. \texorpdfstring{$1$}{} pc), after the electrons have cooled and the emitting region has become optically thin to synchrotron self-absorption due to expansion. 
We present preliminary results of our numerical investigation for the steady-state jet emission and the predicted time lags between \texorpdfstring{$\gamma$}{}-rays and radio during flares. }

\begin{document}




\section{Introduction}\label{sec:1}
Blazars are the most extreme subclass of active galactic nuclei (AGN) having their relativistic jets pointing towards the observer. Their spectral properties are characterized by  non-thermal emission over the entire electromagnetic spectrum, rapid variability, high optical polarization and apparent superluminal motion. One of the characteristic features of blazar jet emission is the shape of its spectral energy distribution (SED), which usually has two components: a low-energy component extending from radio to UV/soft X-rays  and a high-energy component lying between hard X rays and TeV $\texorpdfstring{\gamma}{}$ rays.\\
\indent Multi-wavelength monitoring of blazars offers a unique way of probing the physical conditions in their jets and unveiling the physical processes responsible for their non-thermal variable emission. In recent years, there have been many systematic monitoring programs operating at different energy bands. Their main goal is to investigate the anatomy of blazars by providing high-quality data, probing the rapid variability, and searching for significant correlations between different energy bands, e.g. time lags between radio and $\texorpdfstring{\gamma}{}$-ray emission \cite{P14, HP15, T18}. Fermi-LAT observations  performed over the past 10 years provide crucial spectral and temporal information for hundreds of blazars at energies $>100$ MeV \cite{Acker15}. Being an all-sky $\texorpdfstring{\gamma}{}$-ray monitor  (covering about 20\% of the sky \cite{F09}), Fermi can detect and report on the flaring activity of blazars, thus often triggering follow-up multi-wavelength observing campaigns. Furthermore, HAWC (High Altitude Water Cherenkov) observatory offers the opportunity to study very high-energy (VHE, $E\geq100$ GeV) flares in survey mode, as it scans 2/3 of the entire sky every day \cite{HAWC}. In an attempt to understand the connection of $\texorpdfstring{\gamma}{}$-ray and optical variability in blazars, Robopol (Robotic polarimeter), an optical monitoring program, focuses on
the connection between changes in the optical polarization and $\gamma$-ray flaring activity in blazars \cite{B18}. Multi-year monitoring of blazars at optical and radio frequencies (e.g. BU project \cite{BU17}, F-GAMMA \cite{F16}) can provide information about the structure and the kinematics of the jets at  sub-parsec to tens of parsec scales. Specifically, at radio many programs play a key role in probing the jet morphology  at parcec scales with high resolution radio telescopes and investigating $\gamma$-rays and radio variability correlations (e.g. MOJAVE \cite{LM18}, \cite{HM18}, OVRO \cite{L18, O1, O2}).\\
\indent Despite this multi-year monitoring effort there is still no consensus about the location of the high-energy (X-ray and $\gamma$-ray) emission in blazar jets. In particular, the rapid variability, which is an identifying property of blazars, suggests that the non-thermal emission is typically produced in regions of the jet with size $\ll$ 1 light day  \cite[e.g.][]{aharonian07, aleksic11}. Although electron synchrotron radiation produced in this region can explain the optical-to-X-ray part of the photon spectrum, in most of the cases it cannot account for the jet emission at low frequencies (e.g., $\nu \leq 10^{10}$ Hz); in such compact emitting regions, synchrotron radiation is typically self-absorbed \cite{RL1979}. Yet, blazar jets are detected up to GHz frequencies having a power-law spectrum \cite{G17}. In order to solve this discrepancy, it has been proposed that the radio emission is in most cases not produced in the same region as the $\gamma$-ray photons. The radio-emitting region should be larger in size and less opaque to synchrotron self-absorption \citep{MG85, G85, potter1, M80, M14}.
The results of the various monitoring programs raise several questions about the dynamical and radiative properties of the jets and their variable radio emission \cite{HB2001, KLC2010, GG2011, AT2012}. The localization and the geometry of the emitting source can be estimated by theoretical models in combination with observational data. 
\\
\indent Many studies are based on an idealized model of a conical steady radio jet, while the component which is associated with the variability of the source is related to shock waves that are propagating in the jet \cite{bk79,GM98a,GM98b,katarzynski03}. 
Assuming that electrons are accelerated in such shock fronts and are cooled as they move away from it, one gets that the highest frequency synchrotron component is emitted from a thin layer behind the shock, while the lower frequency component is produced in a larger volume, simply because the cooling time for higher-energy electrons is less than for the lower-energy ones \cite{MG85}.
Furthermore, this scenario predicts time lags between high-frequency and low-frequency light curves and a specific  evolution of the radio spectrum with time.
Building upon these previous studies, \cite{potter1} proposed a model of a ballistic jet with uniform structure in order to investigate the relationship between low and high energy emission \citep[see also][]{katarzynski03,HBS15,RS16}.
In this model, a relativistic electron population is evolved dynamically along the jet, while taking into account the synchrotron and inverse-Compton losses. Then, the synchrotron opacity is self-consistently computed along the line of sight by taking into account the emission at different distances from the jet base. \\ 
\indent The goal of this work is to study the relationship between radio and $\gamma$-ray activity in blazars and to localize their emission using a simplified  framework that allows a wide search of the parameter space. We first construct a \emph{phenomenological} model for radiative transfer (RT) in a conical flow for computing the non-flaring emission of blazars. More specifically, we apply a time-dependent numerical code that has been originally developed to solve the RT equation (RTE) in a spherical geometry \citep{MK95} to compute the emission from a conical expanding outflow \citep[see also][]{ZW16}. To achieve this, we use an expanding spherical volume to solve the RTE \citep[e.g.][]{PM09}, but "disconnect" the dependence  of all other quantities (e.g., magnetic field and particle number density) on the distance  from the black hole. In other words, the functional profiles of the magnetic field and particle injection rate are free and not self-consistently computed. We then present a method for studying flaring episodes by changing one or more parameters for a new blob formed at the base of the jet and following this disturbance as it evolves down the jet. In section \ref{sec:2} we outline the model that we adopt and in section \ref{sec:3} we present our preliminary results. We conclude in section \ref{sec:4} with a summary of our results. 
\section{Model description and numerical approach} \label{sec:2}
In the present study we construct a parametric leptonic model in order to explain the steady-state spectral energy distribution (SED) of blazars by taking into account radiative and adiabatic losses in an expanding volume \cite{S65, V67}.
The emitting region is moving with a highly relativistic speed $v=\beta c$ that corresponds to a Lorentz factor $\Gamma = (1-\beta^2)^{-1/2}$. Then, the Doppler factor of the emitting region is defined $\delta=[\Gamma (1-\beta \cos \theta)]^{-1}$, where $\theta$ is the angle between the jet axis and our line of sight. Relativistic electrons are injected into the source, that is assumed to be spherical with an initial radius $R_0$ as measured in its co-moving frame ( see also end of Section \ref{sec:1}).  The characteristic timescale of the problem is the initial light-crossing time of the blob $t_{cross}=R_0/c$, which translates to an observed variability timescale of $t_{var}=(1+Z) R_0/\delta c$, where $Z$ the redshift; henceforth, we drop all redshift-dependent factors in our analysis. We assume that the emitting region is formed at a distance $z_0$ from the central engine, with $z_0 \gg R_0$, and that its radius increases linearly with time $t$ (as measured in the co-moving frame of the blob) as $R(t)=R_0+u_{exp}t$, where $u_{exp}$ is the expansion velocity ($u_{exp} < c$). The magnetic field strength is parameterized as $B=B_0\left(R_0/R\right)^{s}$, where $B_0$ 
and $s$ are free parameters of the model. The electrons are confined in the emitting region  (i.e., $t_{esc} \rightarrow \infty$) and lose energy due to the adiabatic expansion of the source, synchrotron emission, and inverse Compton scattering. 
\\
\indent In order to calculate the temporal evolution of the electron distribution function, one has to solve two integro-differential equations, each describing the losses/sinks and injection of relativistic electrons and photons in the emitting region \cite{K62}. The kinetic equation of electrons reads:
\begin{equation} \label{eq:2}
\frac{\partial N(\gamma,R)}{\partial R}+\frac{\partial}{\partial \gamma}\left[\left(A_{syn}(\gamma,~R)+A_{ICS}(\gamma,~R)+A_{exp}(\gamma,~R)\right) N(\gamma,R)\right]=Q_e(\gamma,~R), 
\end{equation}
where we used as an independent variable the  co-moving source radius $R$,  which is directly related to co-moving time as described above.  All terms appearing in the above equation have been transformed accordingly (i.e., the rates are measured per unit distance and not per unit time). The terms $A_{syn}, ~A_{ICS}, ~A_{exp}$ are the loss rates for synchrotron emission, inverse Compton scattering, and adiabatic expansion respectively \cite{MK95}. We model the electron injection rate as:
\begin{equation}
Q_e(\gamma,R)= q_e(R) \gamma^{-p} = q_{e_0} \left( \frac{R_0}{R} \right)^{\chi}\gamma^{-p}, \quad \gamma_{min}\leq \gamma\leq\gamma_{max},
\label{eq:Qe}
\end{equation}
where $\chi$  can be positive or negative, and $\gamma_{min}$, $\gamma_{max}$  are, respectively, the minimum and maximum Lorentz factors of the electron distribution. The luminosity of relativistic electrons  injected into the blob of radius $R$ can be derived from eq.~(\ref{eq:Qe}) as $L_e^{inj}=m_{\rm e}c^2u_{exp}\int_{\gamma_{min}}^{\gamma_{ max}} Q_e(\gamma, R) \gamma {\rm{d}}\gamma$.  \\
\indent The frequency below which the synchrotron radiation is absorbed can be derived by the condition $\alpha_{\nu_{ssa}}R \sim 1$ where $\alpha_{\nu_{ssa}}$ is the absorption coefficient (e.g. \cite{RL1979}). The synchrotron self-absorption frequency $\nu_{ssa}$ is a function of radius through its dependence on the magnetic field and electron number density.
In particular, the density of radio-emitting electrons when the blob has radius $R$ can be directly derived from Eq.~\ref{eq:2} as $C_e \approx  t_{cross}u_{exp}V^{-1}\int_{\gamma_1}^{\gamma_2} \int_{R_0}^R Q_e(R^\prime,\gamma)dR^\prime \, d\gamma$, where $V$ is the co-moving volume of the blob and the integration over $\gamma$'s is performed over the relevant range for radio-emitting electrons (the specific values of the integration limits are not relevant for this discussion). We implicitly assumed that the energy loss terms (synchrotron and inverse Compton) for these low-energy electrons  are negligible. Using the above expression, one can then calculate the $\nu_{ssa} $ as follows $\nu_{ssa} (R)\propto \left(R^{\-s (p+2)/2-2}\int_{R_0}^RR^{-\chi}dR\right)^{2/(p+4)}$.  For all the values of $p, s>0$, and $\chi>0$ we explored, we find that $\nu_{ssa}$ decreases for increasing $R$ (in agreement with previous studies). Thus, a blob at small distances from the jet base is optically thick to synchrotron radiation, but it becomes optically thin as it expands and moves to larger distances.  \\
\indent 
To explain the steady-state blazar emission we first assume that blobs with the same initial properties are continuously produced at a distance $z_0$ from the central engine. At any given time an observer will receive emission from a series of blobs with different radii from a wide range of distances. This is \emph{equivalent} to a conical flow with a half-opening angle $\phi =\arctan\left(R_0/z_0\right)$. The distance $z$  traveled by a blob since its "birth", as measured in the black hole's rest frame, is related to its radius $R$ as:  $z=z_0+\beta c (R-R_0) / \Gamma u_{exp}$. 
 Using these assumptions, we integrate along the line of sight the SED in order to reproduce the total steady-state spectrum of the source which is observed. Flaring episodes can be also simulated in this framework, if one blob or a sequence of blobs are injected with different initial properties (e.g. higher value of $B_0$  or increasing magnetic field with radius ($s<0$)) from those producing the steady emission.

\section{Results}\label{sec:3}
\begin{figure}[H]
\centering
\includegraphics[width=0.7\textwidth]{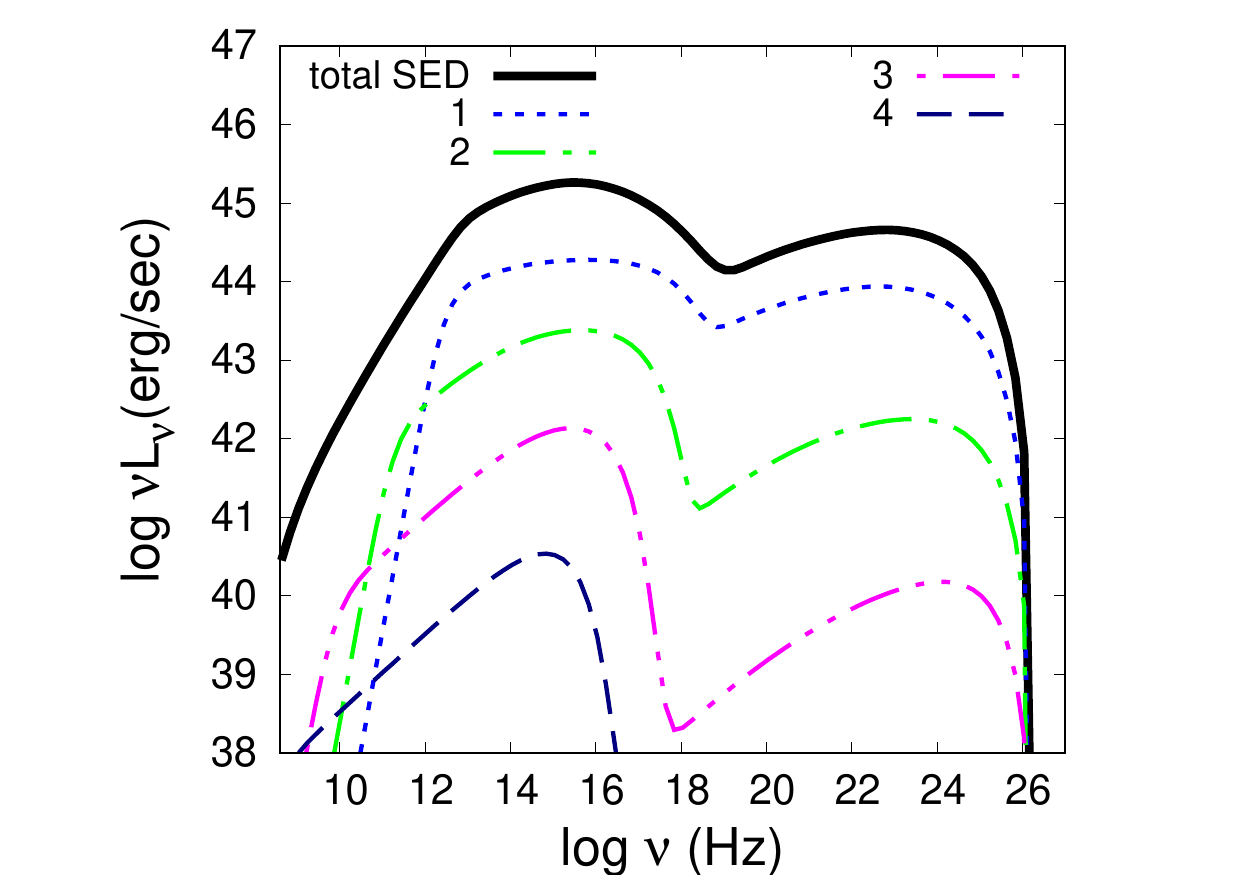}
\caption{Steady-state SED of a fiducial BL Lac source (thick black line), computed by superimposing the emission of $10^4$ blobs that are produced continuously at distance $z_0=10^{-3}$~pc from the central engine. For illustration purposes, we show the spectra of a few indicative blobs ($1: R_1=10^{15.5} ~{\rm cm}, ~z_1=1.6\times 10^{-3} ~{\rm pc}; ~2:R_2=10^{16.5} 
~{\rm cm} ~z_2=7\times10^{-3} ~{\rm pc}; ~3:R_3=10^{17.5} ~{\rm cm}, ~z_3=6\times 10^{-2}~ {\rm pc};~  4:R_4=10^{18.5}~ {\rm cm},~ z_4=z_{final}=6\times 10^{-1} ~{\rm pc}$). All blobs are initialized with the same parameters: $B_0=10$~G, $R_0=10^{15}$~cm, $L_{e_0}^{inj}=10^{42}$ erg s$^{-1}$, $u_{exp}=0.3$ c, $\gamma_{min}=1$, $\gamma_{max}=10^5$, $p=2$, $\Gamma=5$ and $\delta=10$. The magnetic field and electron injection luminosity decrease linearly with radius.}\label{SED}
\end{figure}

In Fig.~\ref{SED} we present an characteristic example for the steady-state emission of a  BL Lac source computed numerically  by superimposing the emission of $10^4$ blobs that are produced continuously at distance $z_0=10^{-3}$~pc from the central engine. We follow the evolution of each blob until it reaches the final distance  $z_{final}=6\times10^{-1}$ pc. The photon spectra emitted by four indicative plasma blobs that are located at different distance $z_i$ from the central engine are overplotted with dashed lines. At the initial time $t_0$, the radius of each blob is $R_0=10^{15}~\rm{cm}$ and its initial magnetic field strength is $B_0=10~\rm{G}$. The bulk Lorentz and Doppler factors are constant  and set to $\Gamma=5$  and $\delta=10$, respectively. Electrons are initially injected with a luminosity $L_{e_0}^{inj}=10^{42}~\rm{erg ~s}^{-1}$ and with a power-law energy distribution (for numerical values, see figure caption). As the blob radius increases with $u_{exp}=0.3~\rm{c}$, the magnetic field strength and electron injection luminosity decrease as $\propto R^{-1}$ (i.e., $s=\chi=1$). The superposition of emission from multiple blobs produces the steady emission of the jet (black thick line).

\begin{figure*}[!htp]   \centering
    \mbox{\includegraphics[width=.48\textwidth,trim={1cm 0.5cm 0 0}]{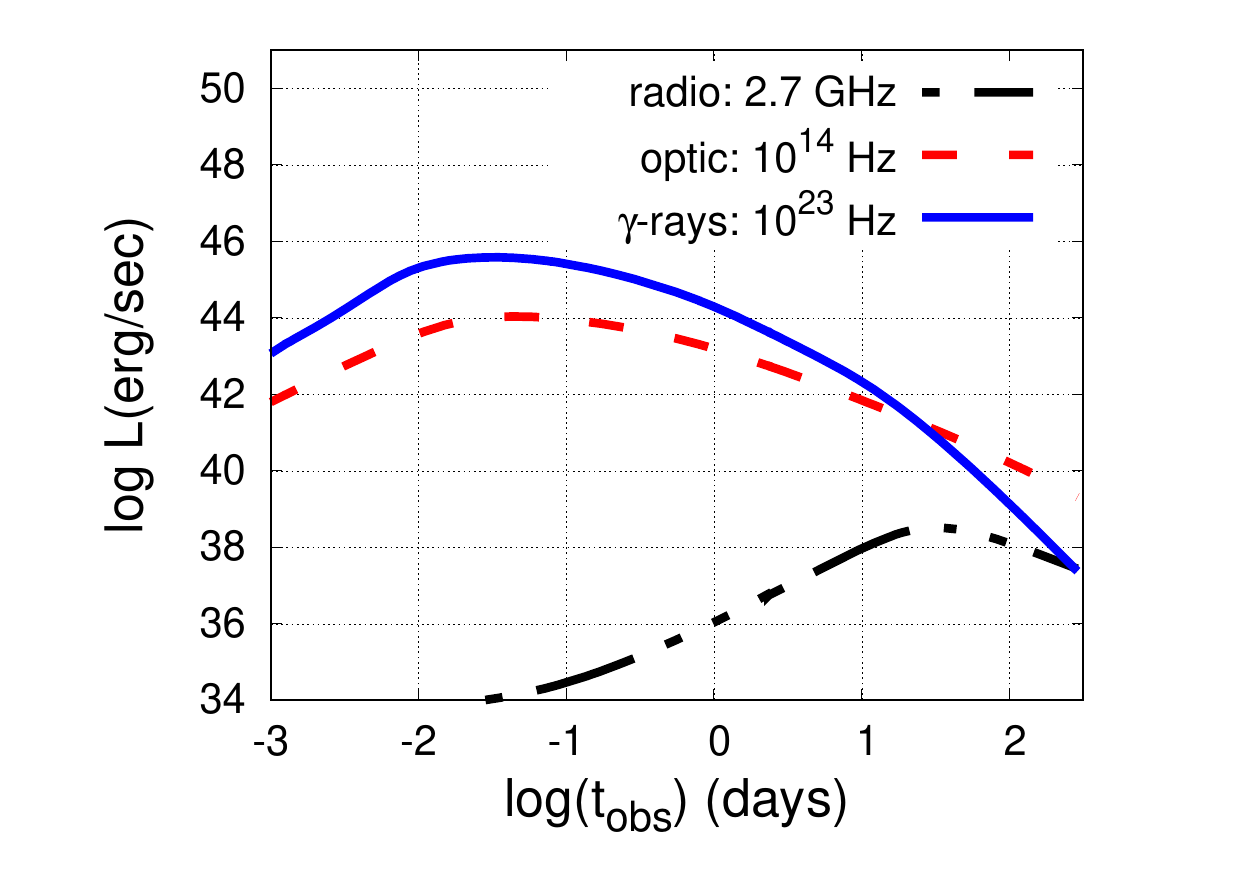}} 
    \mbox{\includegraphics[width=.48\textwidth,trim={1cm 0.5cm 0 0}]{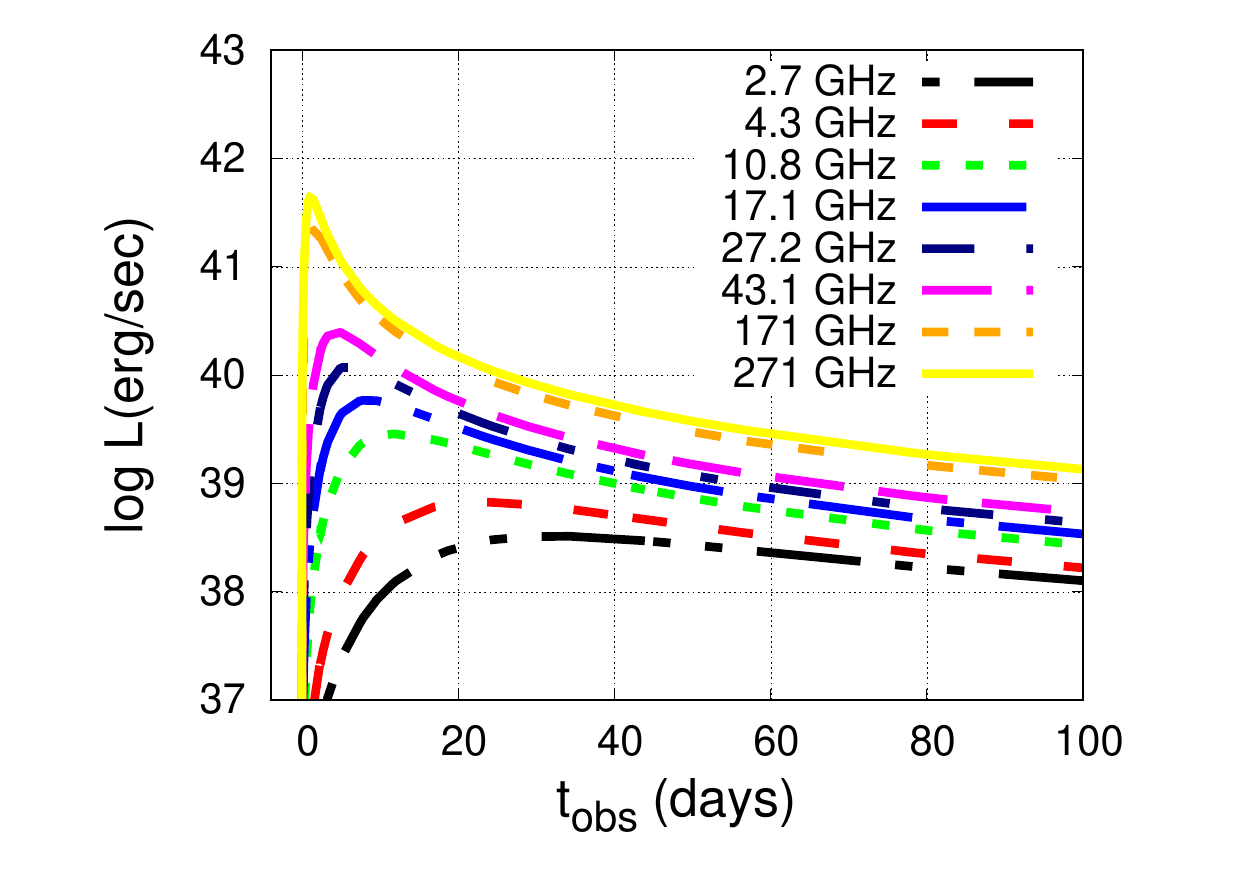}}
 \caption{Multi-wavelength light curves (left panel) and radio light curves (right panel) of an indicative flaring episode produced by  a single blob injected at $z_0=10^{-3}$~pc with the the following properties:
$B_0=1$~G, $R_0=10^{15}$~cm, $L_{e_0}^{inj}=10^{43}$ erg s$^{-1}$, $u_{exp}=0.4$c, $z_{final}$=1~pc, $\gamma_{min}=1$, $\gamma_{max}=10^5$, $p=2$, and $\delta=10$. The magnetic field and electron injection luminosity decrease linearly with radius.}
    \label{LC}
\end{figure*}

\begin{figure*}[!htp] 
\centering
    \includegraphics[width=.315\textwidth, trim=60 0 60 0]{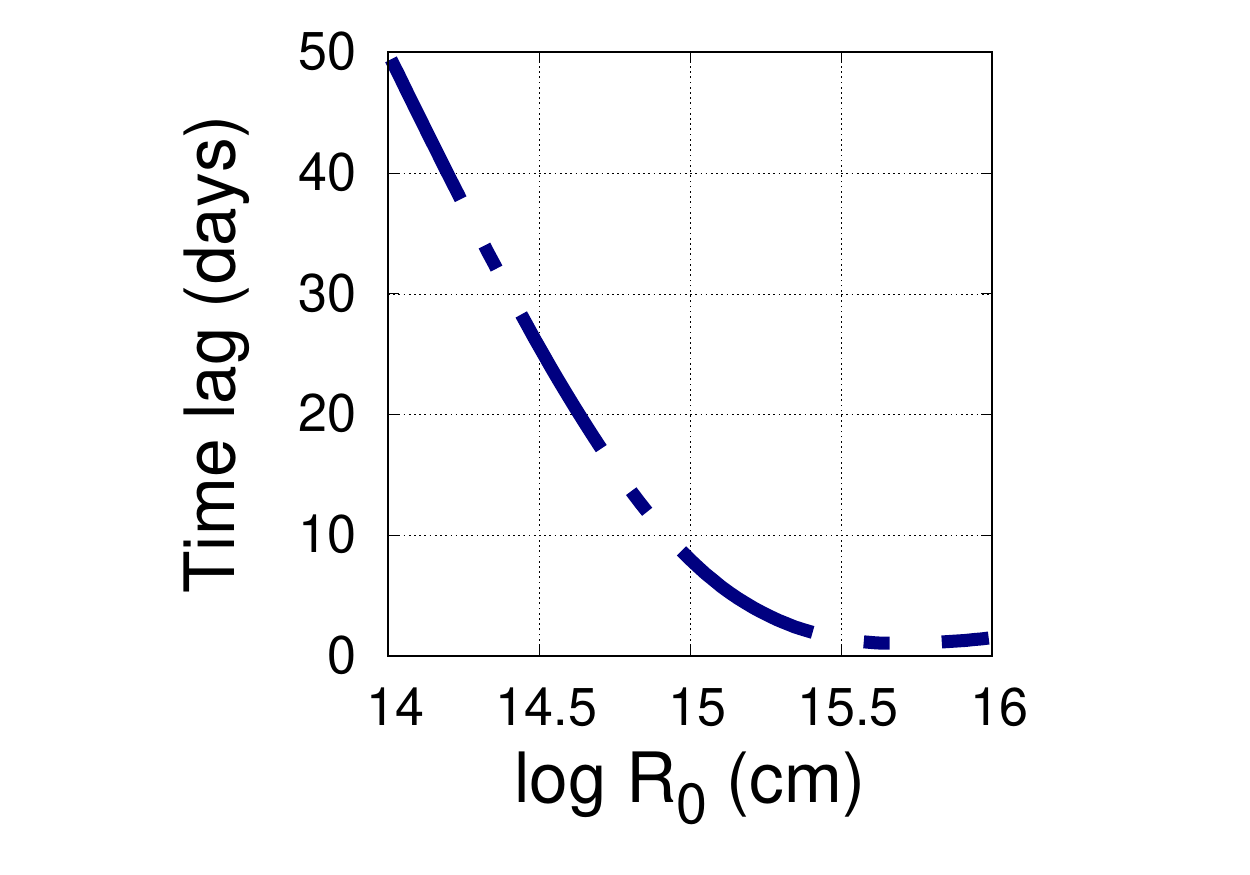}
   \includegraphics[width=.315\textwidth, trim=60 0 60 0]{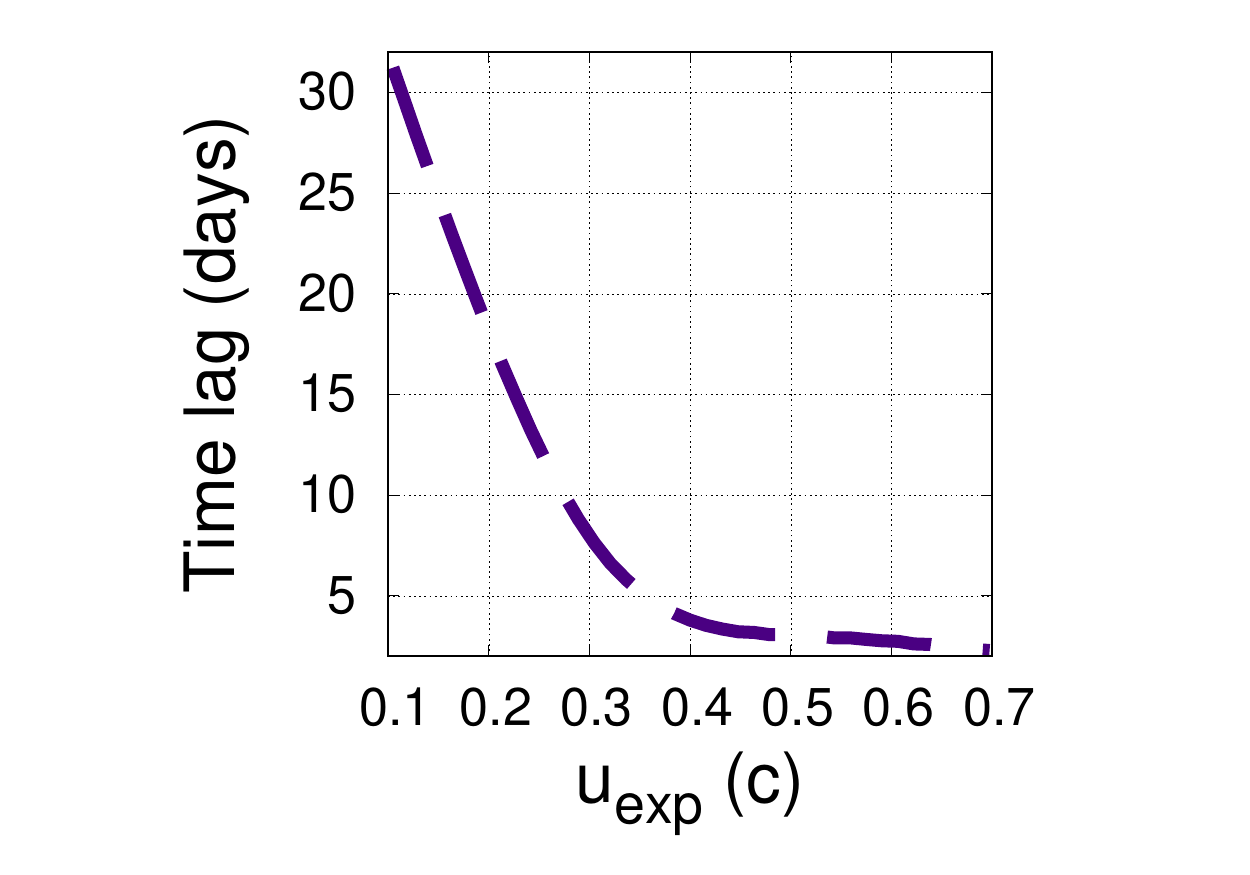}
   \includegraphics[width=.31\textwidth, trim=60 0 60 0]{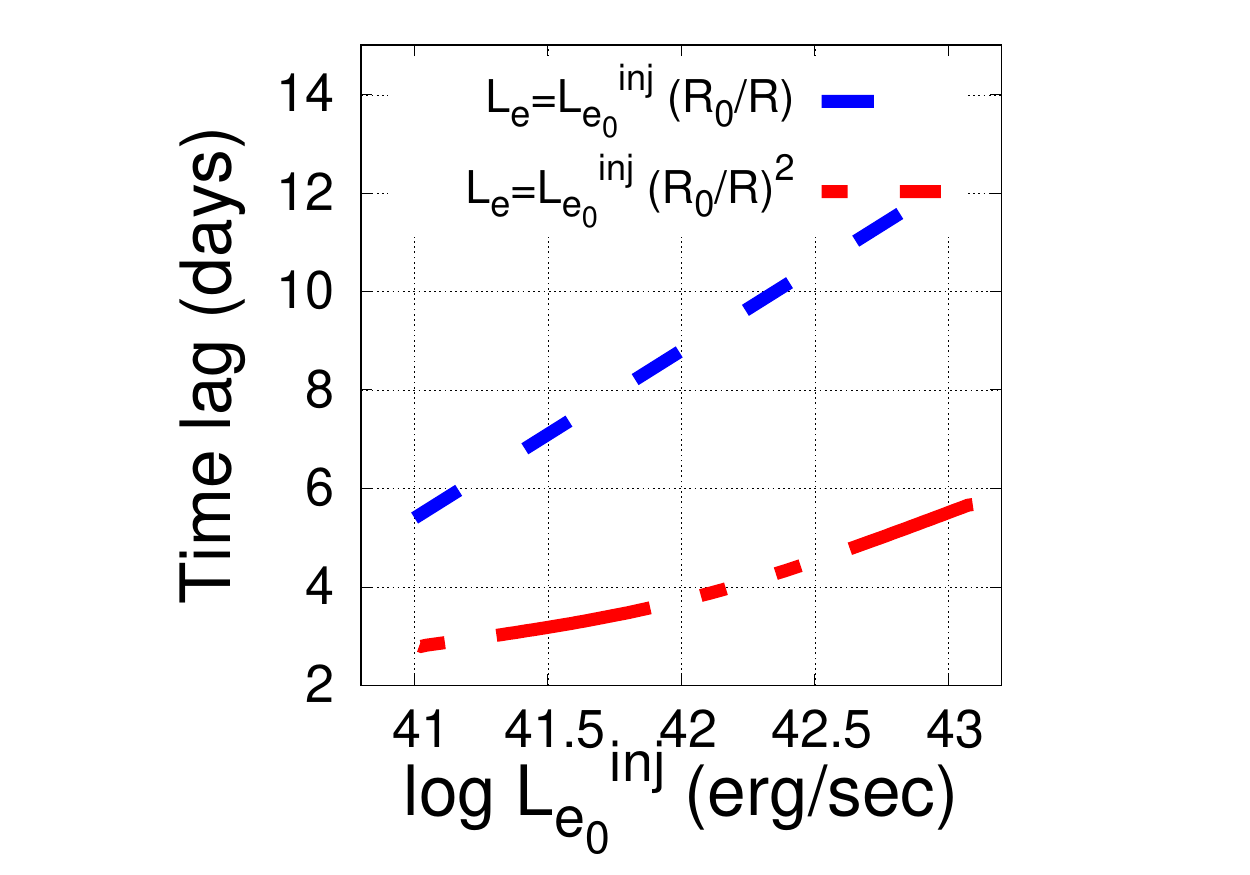}
 \caption{Time lag (in the observer's frame) between the peaks of the  $\gamma$-ray ($10^{23}$ Hz) and GHz light curves as derived by our model for different initial blob parameters (from left to right): blob radius $R_0$, expansion velocity $u_{exp}$, and electron injection luminosity $L_{e_0}^{inj}$ for two different radial dependences marked on the plot.}\label{TL}
\end{figure*}

Figure \ref{LC} depicts the light curves at  different characteristic energy bands ($\gamma$ rays, optical and radio) in the case of a flaring episode.  As can be seen, there is a time lag between the appearance of the $\gamma$ ray emission, which  is produced immediately, and radio, in which there is a delay due to the synchrotron self-absorption. In the co-moving frame, the time lag could be between few hours to few months depending on the choice of the initial parameters. A first effort to study the effects of the initial parameters on the time lags between the $\sim 3$~GHz flux and the 0.4 GeV $\gamma$-ray flux is shown in Fig.~\ref{TL}, where we varied systematically one parameter at each time (e.g. $R_0,~u_{exp}, ~L_{e_0}^{inj}$).

\section{Summary - Discussion}\label{sec:4}
We have presented a method for calculating the steady-state multi-wavelength blazar emission by approximating the jet flow with a superposition of spherical blobs which, after their production at a fixed distance from the black hole, propagate to larger distances while expanding. We follow the evolution of the relativistic electron population of each blob by taking into account radiative energy losses and adiabatic expansion, and compute the resulting photon emission. We then integrate the emission of all blobs that  have propagated at different distances, thus creating an equivalent of a conical outflow, 
to compute the observed spectrum.
This method requires a prescription for the electron injection rate $Q_e$, the magnetic field strength $B$, and the bulk Lorentz factor $\Gamma$ of the flow as function of the distance from the base of the jet, or equivalently as a function of the blob radius $R$. \\
\indent
In the simplest case where $\Gamma$ is constant, $B\propto R^{-1}$, and $Q_e\propto R^{-1}$ we find that the bulk of steady-state emission is produced close to $z_0$, but the SED is severely self-absorbed at $\nu \le 10^{12}$~GHz. Thus, most of the low-frequency steady-state flux emerges at much larger distances, where both the electron number density and the magnetic field are significantly lower due to the expansion of the source (see Fig.~\ref{SED}), in agreement with previous studies (e.g.\cite{V67}). Because of the low photon number density of distant blobs, the absorption of the high-energy photons which are produced at small distances by the more distant ones is negligible. Moreover, the small contribution of the distant blobs to the total steady-state flux suggests that the latter is independent of our choice of the final distance $z_{final}$.  \\
Besides the study of the steady-state emission, our approach can be used to model multi-wavelength flares by changing the initial conditions of a blob, namely by mimicking a perturbation in the flow properties  at an arbitrary distance from $z_0$. The perturbation  (or "active" blob) then moves down to the jet and results in a different SED. Figures \ref{LC} and \ref{TL} show the results of the evolution of such perturbation parametrized by a single blob as it expands. We predict zero or positive time lags, i.e. the $\gamma$-rays come first and the radio follow. Our results agree with most of the cases that are observed (e.g. \cite{HP15}).
Alternatively, a flare may be produced by  re-acceleration of electrons at a large distance from $z_0$. Our approach can also be applied to this scenario by changing the properties (e.g., electron injection rate and magnetic field strength) of a blob after it has reached a certain distance.  In the present work we chose to explore cases in which the magnetic field strength and electron injection rate of all blobs (those producing the steady and flaring emission) decrease with distance from the central engine, as we expect in a steady-state jet. In a follow-up extended paper, we will investigate cases with increasing $Q_e$ and $B$ when studying flaring episodes. 

So far, we have considered a unique and constant Lorentz factor for all the blobs. Nevertheless, our approach can be extended to take into account changes in the bulk Lorentz factor with distance to study the effects of bulk flow acceleration (or deceleration) on both the steady-state and flaring emission  \citep[e.g.][]{GM98a, GM98b}. Although we have presented results for BL Lac objects, we can expand our method by including a varying contribution of external photon fields as the blobs propagate away from the central engine.\\
\indent In conclusion, we have presented a time-dependent numerical model for computing both the steady-state and flaring multi-wavelength emission from blazars.  This model has new ingredients in comparison to previous works:  it can take into account synchrotron self-Compton losses, which could be important in some flaring cases,  it can be used to investigate flaring events by changing many of the parameters of the blob, and it can treat particle acceleration during flares by including the appropriate terms in the electron equation. Thanks to its parametric nature, the proposed framework allows a wide search of the parameter space  and a comparison with previous physical jet models \citep[e.g.][]{GM98a, katarzynski03}, which will be the topic of a future publication.

\vspace{6pt} 

\section*{Aknowledgments}
The authors would like to thank the three anonymous reviewers for their constructive criticism and comments. SB: This research is co-financed by Greece and the European Union (European Social Fund-ESF) through the Operational Programme «Human Resources Development, Education and Lifelong Learning» in the context of the project “Strengthening Human Resources Research Potential via Doctorate Research” (MIS-5000432), implemented by the State Scholarships
Foundation (IKY).
MP acknowledges support from the L. Spitzer Postdoctoral Fellowship and NASA grant 80NSSC18K1745.




\end{document}